\documentclass[a4paper,12pt]{article}
\usepackage{epsfig,rotating}
\newcommand{\Zo}{\mbox{$\mathrm{Z^0}$}}
\newcommand{\WW}{\mbox{$\mathrm{W}^+\mathrm{W}^-$}}
\newcommand{\Hp}{\mbox{$\mathrm{H}^{+}$}}
\newcommand{\Hm}{\mbox{$\mathrm{H}^{-}$}}
\newcommand{\Hpm}{\mbox{$\mathrm{H}^{\pm}$}}
\newcommand{\ee}{\mbox{$\mathrm{e}^{+}\mathrm{e}^{-}$}}
\newcommand{\pb}{\mbox{$\mathrm{pb}^{-1}$}}

\newcommand{\fbp}{\mbox{$\mathrm{fb}^{-1}.$}}
\newcommand{\ra}{\mbox{$\rightarrow$}}
\newcommand{\tp}{\mbox{$\tau^+$}}
\newcommand{\tm}{\mbox{$\tau^-$}}

\newcommand{\csbar}{\mbox{$\mathrm{c} \bar{\mathrm{s}}$}}
\newcommand{\cbars}{\mbox{$\bar{\mathrm{c}} \mathrm{s}$}}

\newcommand{\nubar}{\mbox{$\bar{\nu}$}}
\newcommand{\mHp}{\mbox{$m_{\mathrm{H}^{\pm}}$}}
\newcommand{\mA}{\mbox{$m_{\mathrm{A}}$}}
\newcommand{\mHpm}{\mbox{$m_{\mathrm{H}^{\pm}}$}}

\newcommand{\Gcs}{\mbox{${\rm GeV}/c^2$}}
\newcommand{\Mcs}{\mbox{${\rm MeV}/c^2$}}

\newcommand{\tanb}{\mbox{$\tan \beta$}}

\newcommand{\HWAWA}{\mathrm{W}^*\!\mathrm{A}\mathrm{W}^*\!\mathrm{A}}
\newcommand{\HWATN}{\mbox{$\mathrm{W^{*}A}\tau^{-}\bar{\nu_{\tau}}$}}
\newcommand{\HWACS}{\mbox{$\mathrm{W^{*}A}\bar{\mathrm{c}}\mathrm{s}$}}
\newcommand{\wa}{\mathrm{W}^*\mathrm{A}}
\newcommand{\wn}{\mathrm{W}^*\!\Phi}
\newcommand{\hphm}{\mathrm{H}^+\mathrm{H}^-}
\begin{document}
\begin{titlepage}
\begin{center}
\vspace{-0.5cm}
{\Large EUROPEAN ORGANIZATION FOR NUCLEAR RESEARCH}
\end{center}
\begin{flushright}
  CERN-PH-EP-2012-369 \\
  December 19, 2012 
\end{flushright}
\begin{center}{\Large \bf Search for Charged Higgs bosons: \\
Combined Results Using LEP data}
 \bigskip
\end{center}
\begin{center}
      {\large  ALEPH, DELPHI, L3 and OPAL Collaborations\\
      The LEP working group for Higgs boson searches\footnote{The authors are listed in Refs.[3,4,6,7,9].}}
\end{center}
\bigskip
\begin{center}{\Large  Abstract}\end{center}
The four LEP collaborations, ALEPH, DELPHI, L3 and OPAL,
have searched for pair-produced charged Higgs bosons 
in the framework of Two Higgs Doublet Models (2HDMs). The data of the four 
experiments have been statistically combined. The results are interpreted 
within the 2HDM for Type I and Type II benchmark scenarios. 
No statistically significant excess has been observed when compared 
to the Standard Model background prediction, and the combined LEP data exclude 
large regions of the model parameter space. 
Charged Higgs bosons with mass below 80 \Gcs\ (Type II scenario)
or 72.5 \Gcs\ (Type I scenario, for pseudo-scalar masses above 12 \Gcs) 
are excluded at the 95\% confidence level.

\bigskip
\begin{center}
{\it Accepted by Eur. Phys. Journal C}
\end{center}
\end{titlepage}


\section{Introduction}
\label{intro}
A charged Higgs boson appears in many extensions of the Higgs sector beyond
the Standard Model (SM). Indeed, its discovery would signal unambigously that
the Higgs-like particle recently discovered at LHC~\cite{higgslike}
is not the SM neutral Higgs
boson. It is thus of great interest to search for a charged Higgs boson.

More than twelve years after the end of data-taking at LEP, it is 
important to add to the LEP legacy the outcome of the searches for a charged 
Higgs boson. In fact, the charged Higgs boson searches at a lepton collider
are significantly less model-dependent than the corresponding searches at
hadron colliders, due to the very simple production mechanism. 

Since the previous communication by the LEP working group for Higgs 
boson searches (LEPHWG) on charged Higgs boson, 
in 2001~\cite{previous}, 
the LEP experiments have published their final results on these searches 
and have, in some cases, also added searches for new final states not 
previously considered.
The four LEP collaborations have searched for charged Higgs bosons in the
framework of Two Higgs Doublet Models (2HDMs). Based on the final results
obtained by ALEPH~\cite{aleph}, DELPHI~[4,5], L3~\cite{l3} and
OPAL~[7,8], the LEPHWG has performed 
a statistical combination of the data taken at centre-of-mass energies,
$\sqrt{s}$, from 183~GeV to 209~GeV.
The total luminosity used in this combination is 2.6~\fbp 

In 2HDMs~\cite{2HDM}, there are five physical Higgs bosons: 
the CP-even h and H, the CP-odd
A and the charged Higgs bosons, \Hpm. The charged Higgs couplings to 
the photon and the Z boson are completely specified
in terms of the electric charge and the weak mixing angle, $\theta_W$,
and therefore, at tree level, the production cross-section depends only on 
the charged Higgs boson mass. Higgs bosons couple proportionally to the 
particle mass and therefore decay
preferentially to heavy particles, but the precise 
branching ratios may vary significantly depending on the model.
Two scenarios are considered in this paper.
The first one effectively allows the charged Higgs boson to decay to fermions 
only, which is the case
in type II 2HDM for not too small values of \mA\ 
(the neutral CP-odd A boson mass) or $\tan\beta$
(the ratio of the two Higgs doublet vacuum expectation values). 
In this model the isospin $+\frac{1}{2}$ fermion-couplings to the charged 
Higgs boson are  proportional to $1/\tan\beta$, while the 
isospin $-\frac{1}{2}$ fermion-couplings are proportional to $\tan\beta$.
This scenario is treated in Section 2.
In the second scenario, type I 2HDM, all fermions couple 
proportionally to
$1/\tan\beta$. Consequently, the second scenario effectively allows 
the charged Higgs boson to also decay into gauge 
(possibly off-shell) bosons and Higgs bosons (see Section 3).

Pair-production of charged Higgs bosons via s-channel exchange of a
\Zo\ boson would modify the decay width of the \Zo\ boson. Therefore
electroweak precision measurements set indirect bounds on the mass of the 
charged Higgs boson regardless of its decay branching ratios. 
The difference between the measured decay width of the 
$\Zo$ ($\Gamma_{\mathrm{Z}}$) and the prediction from the SM
sets a limit on any non-standard (non SM) contribution to $\Zo$ decay.
The $\Zo$ decay width has been measured precisely during 
the first phase of LEP (LEP-1).
The final LEP result~\cite{lepew} set the limit 
$\Gamma_{\rm non SM}<2.9$ MeV$/c^2$ (95\% C.L.),
which translates to $\mHp >$ 39.6 GeV$/c^2$ (95\% C.L.).
Direct searches during the LEP-1 period 
set a lower bound for the charged Higgs boson mass 
at 44.1 GeV$/c^2$ at 95\% C.L. for type II 2HDM [12-15].
The combination in this paper is performed for charged 
Higgs boson masses of 43 GeV$/c^2$ or larger, since the region below 43 \Gcs\ 
has been covered by individual experiments. 

For this combination of data, the cross-sections (and branching ratios   
for type II 2HDM) are calculated within the HZHA program package~\cite{hzha}, 
and the branching ratios of the charged Higgs boson in type I 2HDM are taken 
from Ref.~\cite{akeroyd}. 

The input from the four experiments [3-8] which is used in the combination 
procedure is provided on a channel-by-channel basis. The word ``channel'' 
designates any 
subset of the data where a Higgs boson search has been carried out. 
Table~\ref{catalog} shows a summary of all channels available for this 
combination. It amounts to 22 channels from ALEPH, 43 from DELPHI, 
12 from L3 and 45 from OPAL.

Each experiment generated and simulated the detailed detector response in
Monte Carlo event samples for 
the Higgs signal and the various background processes, at centre-of-mass
energies of 183,
189, 192, 196, 200, 202, 204, 206, 208 and 209~GeV to estimate background and
signal contributions in the data collected between 1997 and 2000.
Particular care has been taken when simulating the four-fermion background,
especially from W-pair background, using the most advanced codes available at 
that time.
ALEPH used KORALW~\cite{koralw} as the generator and RACOONWW~\cite{racoon} 
and YFSWW~\cite{yfsww} for the cross-section calculation, 
while DELPHI used WPHACT~\cite{wphact}, L3 YFSWW 
and OPAL GRC4F~\cite{grc4f} and KORALW. Other generators were used for
systematic studies.
Furthermore, each of the four experiments used different values for the W mass 
in these background simulations (respectively 
80.45, 80.40, 80.356 and 80.33 \Gcs~for ALEPH, DELPHI, L3 and OPAL), 
while the LEP combined measured value is $80.376 \pm 0.033$
\Gcs~\cite{wmass}, 
thus introducing an additionnal source of systematic uncertainty due to the
broadening of the W peak when adding the four backgound simulations.

\begin{table}
\caption{\small\it Overview of the searches for charged Higgs bosons performed
by the four LEP experiments, whose results are used in this combination.
Where relevant, \mA\ varies from 2$m_b$ to $m_{\Hpm}$. Each experiment analysed
typically around 650 \pb of data.}
\label{catalog}
\begin{center}
\begin{tabular}{llll}
\hline\noalign{\smallskip}
Expt & Final state & $\sqrt{s}$ & $m_{\Hpm}$\\
(Ref.)  &    &                 & range \\
        &    & (GeV) & (GeV/$c^2$) \\
\noalign{\smallskip}\hline\noalign{\smallskip}
ALEPH  & ${\mathrm H}^+{\mathrm H}^- \rightarrow \csbar\cbars $ & 
189 - 209 & 45 - 100 \\
\cite{aleph}   & ${\mathrm H}^+{\mathrm H}^- \rightarrow \csbar\tau\nu $ & 
189 - 209 & 55 - 100 \\
       & ${\mathrm H}^+{\mathrm H}^- \rightarrow \tau\nu\tau\nu $ & 
189 - 209 & 45 - 100 \\
\noalign{\smallskip}\hline
DELPHI & ${\mathrm H}^+{\mathrm H}^- \rightarrow \csbar\cbars $ & 
183 - 209 & 40 - 100 \\
\cite{delphi},\cite{del183}&${\mathrm H}^+{\mathrm H}^- \rightarrow 
\csbar\tau\nu $ & 
183 - 209 & 40 - 100 \\
       & ${\mathrm H}^+{\mathrm H}^- \rightarrow \tau\nu\tau\nu $ & 
183 - 209 & 40 - 100 \\
       & ${\mathrm H}^+{\mathrm H}^- \rightarrow {\mathrm W}^*{\mathrm A}
\tau\nu $ & 
189 - 209 & 40 - 100 \\
       & ${\mathrm H}^+{\mathrm H}^- \rightarrow {\mathrm W}^*{\mathrm A}
{\mathrm W}^*{\mathrm A} $ & 
189 - 209 & 40 - 100 \\
\noalign{\smallskip}\hline
L3     & ${\mathrm H}^+{\mathrm H}^- \rightarrow \csbar\cbars $ & 
183 - 209 & 50 - 100 \\
\cite{l3}  & ${\mathrm H}^+{\mathrm H}^- \rightarrow \csbar\tau\nu $ & 
183 - 209 & 50 - 100 \\
       & ${\mathrm H}^+{\mathrm H}^- \rightarrow \tau\nu\tau\nu $ & 
183 - 209 & 50 - 100 \\
\noalign{\smallskip}\hline
OPAL   & ${\mathrm H}^+{\mathrm H}^- \rightarrow \csbar\cbars $ & 
183 - 209 & 40 - 100 \\
\cite{opal},\cite{opa183}&${\mathrm H}^+{\mathrm H}^- \rightarrow 
\csbar\tau\nu $ & 
183 - 209 & 40 - 100 \\
       & ${\mathrm H}^+{\mathrm H}^- \rightarrow \tau\nu\tau\nu $ & 
183 - 209 & 45 - 100 \\
       & ${\mathrm H}^+{\mathrm H}^- \rightarrow {\mathrm W}^*{\mathrm A}
\tau\nu $ & 
189 - 209 & 40 - 95 \\
       & ${\mathrm H}^+{\mathrm H}^- \rightarrow {\mathrm W}^*{\mathrm A}
{\mathrm W}^*{\mathrm A} $ & 
189 - 209 & 40 - 95 \\
\noalign{\smallskip}\hline
\end{tabular}
\end{center}
\end{table}

The statistical procedure adopted for the combination of the data 
and the precise definitions of the confidence levels 
$CL_b,~CL_s,~CL_{s+b}$ by which the search results are expressed, 
have been described previously~\cite{adlo-cernep}.
The main sources of systematic uncertainty affecting the signal and 
background predictions are included, using an
extension of the method of Cousins and Highland~\cite{cousins-highland} 
where the p-values are averaged over a large ensemble of 
Monte Carlo experiments. The correlations between search 
channels, LEP collision energies and individual experiments have not been 
taken into account, but these correlations are estimated to have only small
effects, about 500 \Mcs , to the final results.

\section{Combined searches in the framework of type II 2HDM}
\label{sec:2}
In type II 2HDM, one Higgs doublet couples to up-type 
fermions and the other to down-type fermions. The Higgs sector of the
Minimal Supersymmetric Standard Model (MSSM) is a particular case
of such models. In the MSSM, at tree-level,
the \Hpm\ is constrained to be heavier than the W boson and 
the radiative corrections to the charged Higgs mass are positive, except  
for very specific parameter choices. 
Thus, experimentally finding evidence of
a charged Higgs boson with mass below the W boson mass
would set very strong constraints on the MSSM parameters.
However, in the following we will concentrate on the general type~II 2HDM 
without any supersymmetric assumptions. Results on the search for neutral 
MSSM Higgs bosons can be found in~\cite{MSSM}.

For the charged Higgs masses accessible at LEP energies, 
the decays into $\tau^+\nu_{\tau}$ and $\csbar$ (and their charge conjugates)
are expected to dominate.
The searches are carried out under the assumption that 
the two decays \Hp\ra\csbar\ and \Hp\ra\tp$\nu$ exhaust the \Hp\ decay
width, but the relative branching ratio is free. 
This assumption is valid as long as
\mA\ is larger than 60 \Gcs\ (MSSM case) or $\tan\beta$ is larger than a few 
units. Thus, the 
searches encompass the following $\hphm$ final states: (\csbar)(\cbars),
(\tp$\nu$)(\tm\nubar) and the mixed mode 
(\csbar)(\tm\nubar) or (\cbars)(\tp$\nu$).
The combined search results are presented as a function of the branching 
ratio Br(\Hp\ra\tp$\nu$).

Details of the searches done by the individual experiments 
can be found in Refs.~[3-8].
Two features in these analyses are worth noting: the main background is
W pair production, which is partly irreducible, and the
reconstructed mass
is one of the discriminant variables used in the final hypothesis testing
in the two channels where this is relevant (mixed and hadronic channels). 
The results from the four LEP experiments are summarised in 
Table~\ref{table-ch-input}, 
together with the 95\% C.L. observed and median expected lower limits on the 
charged Higgs boson mass. The mass limits are quoted separately for 
Br(\Hp\ra\tp$\nu$) = 0, 1, and independently of the charged Higgs decay.

\begin{table*} [hbtp]
\begin{center}
\caption{\small\it Individual search results for the
\ee\ra\Hp\Hm\ fermionic final states. 
All limits are given at the 95\% C.L.
The OPAL selection  is mass-dependent; 
the numbers of events given here are for \mHpm = {\rm 80}~\Gcs.}
\label{table-ch-input}
\begin{tabular}{lcccc}
\noalign{\smallskip}\hline
Experiment        & ALEPH [3] & DELPHI [4] & L3 [6] & OPAL [7] \\
\noalign{\smallskip}\hline
Total Int. luminosity (\pb)  &630   & 620      & 685    &670   \\
\phantom{.....}Final states & \multicolumn{4}{c}{Number of expected/observed  
events~~~~}\\
\phantom{..........}(\csbar)(\cbars)  &2806.0/2742&2179.3/2179&2473.8/2578&
1501.4/1471 \\
\phantom{..........}(\csbar)(\tm\nubar)    &289.3/280&1122.8/1129&494.5/470&
526.3/569 \\
\phantom{..........}(\tp$\nu$)(\tm\nubar) &39.8/45& 73.6/ 66&149.8/147 &
1103.4/1110\\
\phantom{..........}Sum of all channels   &3135.1/3067&3375.7/3374&3118.1/3195&
3131.1/3150\\
\noalign{\smallskip}\hline
Mass limits in \Gcs  &  &  &  &  \\
Expected(median)/ observed limit    &  &  &  &  \\
\phantom{.....}Br(\Hp\ra\tp$\nu$)=0  & 78.2/80.4 & 77.7/77.8 & 76.8/76.6 & 77.2/76.5 \\
\phantom{.....}Br(\Hp\ra\tp$\nu$)=1  & 89.2/87.8 & 88.9/90.1 & 84.3/83.7 & 89.2/91.3 \\
\phantom{.....}any Br(\Hp\ra\tp$\nu$)& 77.1/79.3 & 76.3/74.4 & 75.7/76.4 & 75.6/76.3 \\
\noalign{\smallskip}\hline
\end{tabular}
\end{center}
\end{table*}

In a first step, the statistical combination software was run separately on 
the data provided by the four collaborations and the results 
compared to the published results of each.
The differences between this check and the published results (of the order of
$\pm 200$ \Mcs\ in the limits) reflect the
differences between the statistical methods used by the four collaborations.
The biggest difference has been found for the expected limit from OPAL at
Br(\Hp\ra\tp$\nu$)= 1 and amounts to 600 \Mcs. This difference is compatible
in size with the estimated effect of about 500 \Mcs\ of
not taking into account the correlations between systematic 
uncertainties. All mass limits have thus
been rounded down to the nearest half a \Gcs.  

Combining the results from the four experiments, a scan in the branching 
ratio Br(\Hp\ra\tp$\nu$) versus charged Higgs boson mass plane has been 
performed, and the limit-setting procedure was repeated for each scan point.
This two-dimensional scan was performed with the following ranges
and steps: \mHpm\ from 43 to 95 \Gcs\ with 1 \Gcs\ steps, and 
Br(\Hp\ra\tp$\nu$) from 0 to 1 with 0.05 steps.

\begin{figure}[htb]
\begin{center}
\includegraphics[width=0.90\columnwidth]{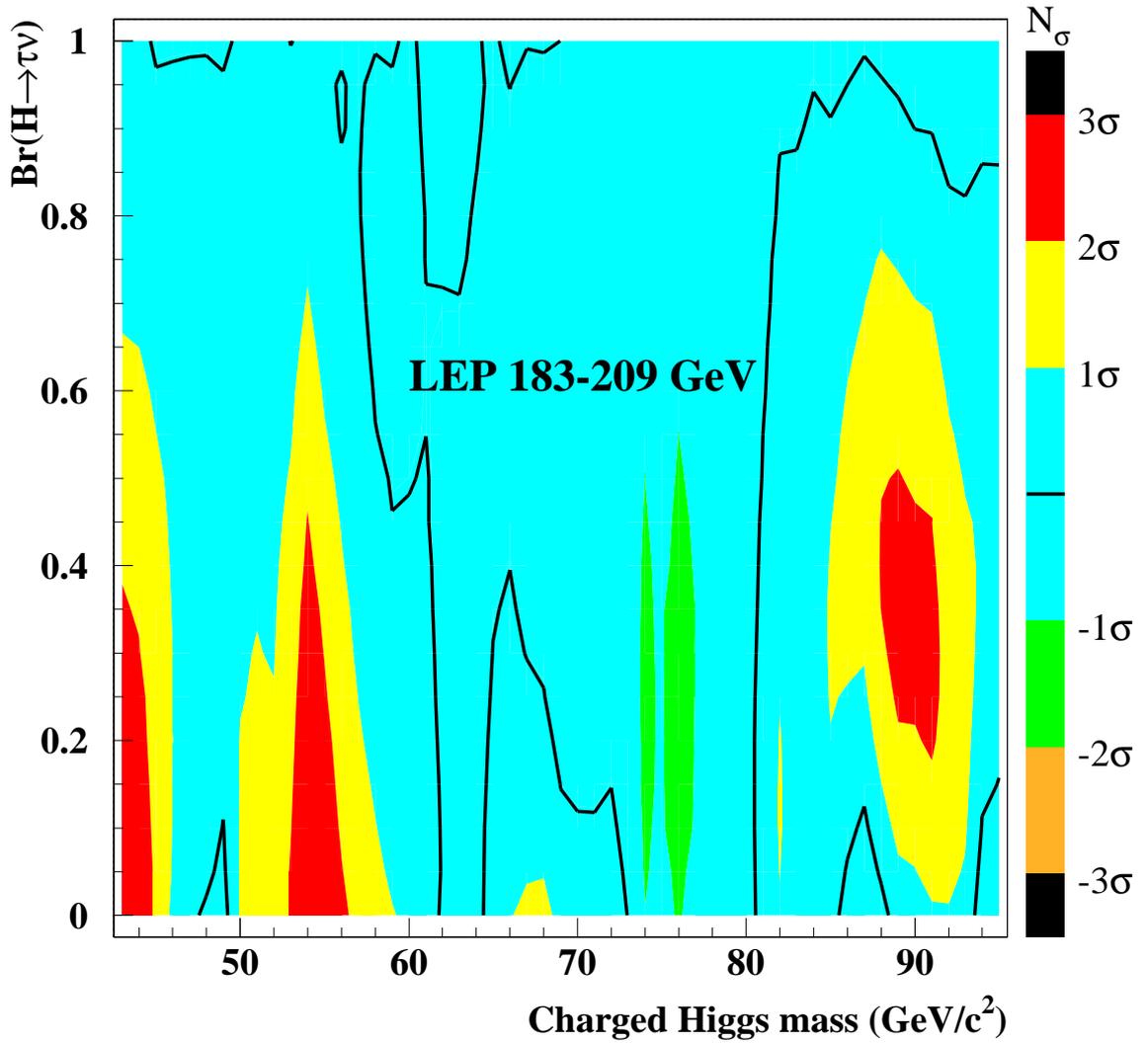}
\caption[]{\small \it Type II 2HDM: contours based on the observed p-values 
$CL_b$ as a function of \mHpm\ and the branching ratio
Br(\Hp\ra\tp$\nu$), indicating the statistical significance, N$_{\sigma}$,
of local departures from the background expectation. The black solid line 
indicates the change of sign of this significance, i.e. where
there is a transition from excess to deficit.}
\label{charged-clb}
\end{center}
\end{figure}

Figure~\ref{charged-clb} shows the observed background confidence level 
$CL_b$ as a function of \mHpm\ and Br(\Hp\ra\tp$\nu$). 
The observed confidence level is everywhere within 
$\pm 2\sigma$ of the background prediction, except 
for three small regions, as shown in Figure 1.
Such regions result from the combination of 
small excesses, as compared to the background expectation, observed by two 
or three experiments. The first two mass regions, around 43 and 55 \Gcs, 
are due to a slight excess of data 
in the hadronic channel, and the third one, around 90 \Gcs, arises from
an excess of events in the mixed channel. Table~\ref{hotspots} gives the
combined $CL_b$ together with the values from each experiment for these
three domains (the values chosen for Br(\Hp\ra\tp$\nu$) are given in the
second column).

\begin{table*} [hbtp]
\begin{center}
\caption{\small\it Combined and individual $CL_b$ values for the three
mass points with a deviation from expectation larger than 2 $\sigma$. All
values, obtained with the statistical procedure of the overall
combination, compare well to those published by the experiments.
(*) ALEPH and L3 did not provide inputs for this mass.}
\label{hotspots}
\begin{tabular}{ccccccc}
\noalign{\smallskip}\hline
$\mHp$&Br(\Hp\ra\tp$\nu$)&combined& ALEPH & DELPHI & L3  & OPAL \\
(\Gcs)  &   &  $CL_b$   & $CL_b$  & $CL_b$  & $CL_b$ & $CL_b$   \\
\noalign{\smallskip}\hline
43.    &  0.0      & 0.998   & (*)    & 0.99   & (*)   & 0.96  \\
\noalign{\smallskip}\hline
55.    &  0.0      & 0.997   & 0.75   & 0.96   & 0.96  & 0.94  \\
\noalign{\smallskip}\hline
89.    &  0.35     & 0.988   & 0.98   & 0.63   & 0.88  & 0.80  \\
\noalign{\smallskip}\hline
\end{tabular}
\end{center}
\end{table*}

\begin{figure}[htb]
\begin{center}
\includegraphics[width=0.50\columnwidth]{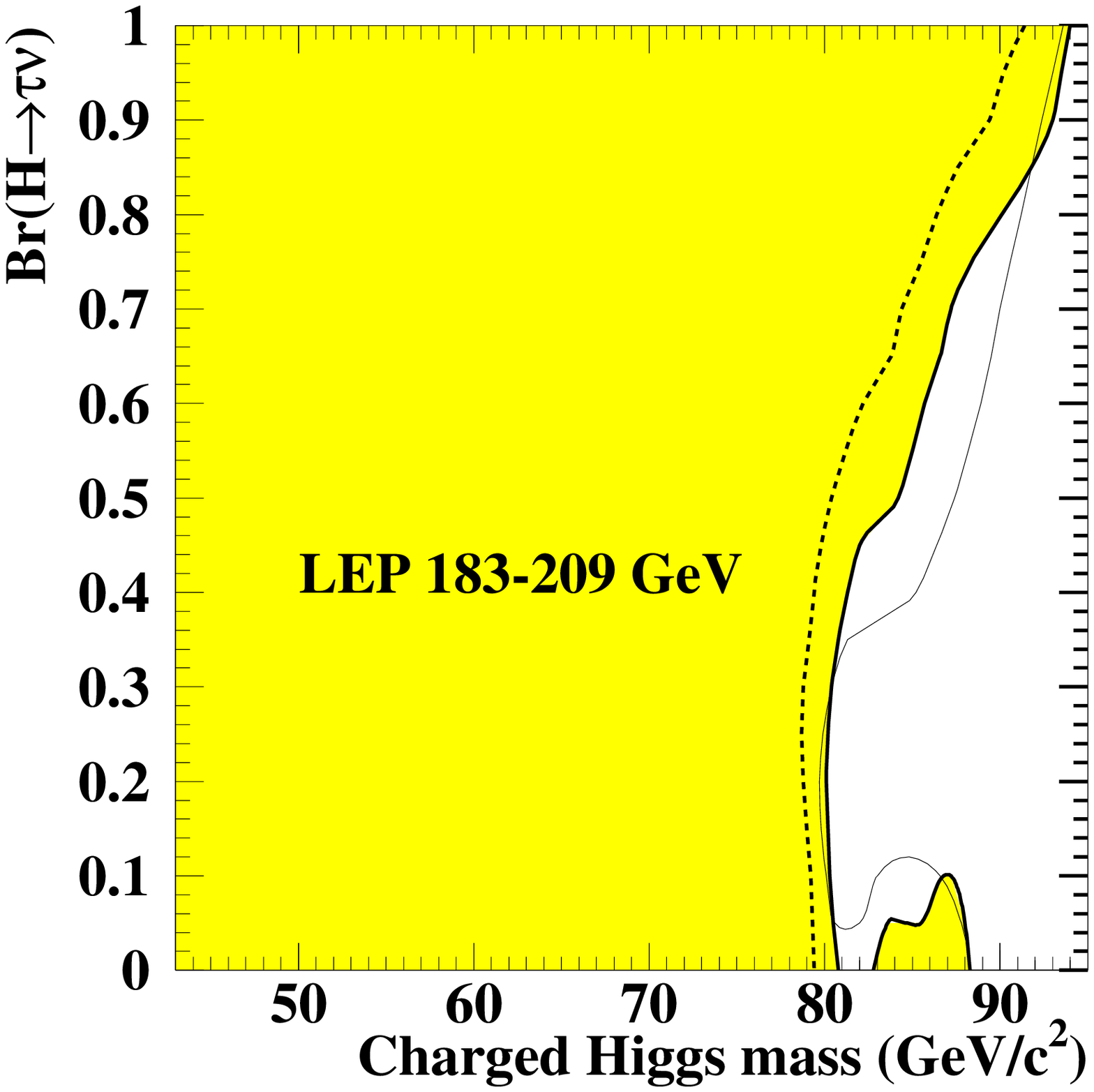}
\caption[]{\small \it Type II 2HDM: 
excluded regions in the Br(\Hp\ra\tp$\nu$) vs \mHpm\ plane,
based on the combined data collected by the four LEP experiments 
at centre-of-mass 
energies from 183 to 209~GeV. The shaded area is excluded at the 95\% or 
higher C.L. 
The expected exclusion limit (at the 95\% C.L.) is indicated by the thin solid 
line and the thick dotted line inside the shaded area is the observed limit 
at the 99.7\% C.L.}
\label{charged-limit}
\includegraphics[width=0.50\columnwidth]{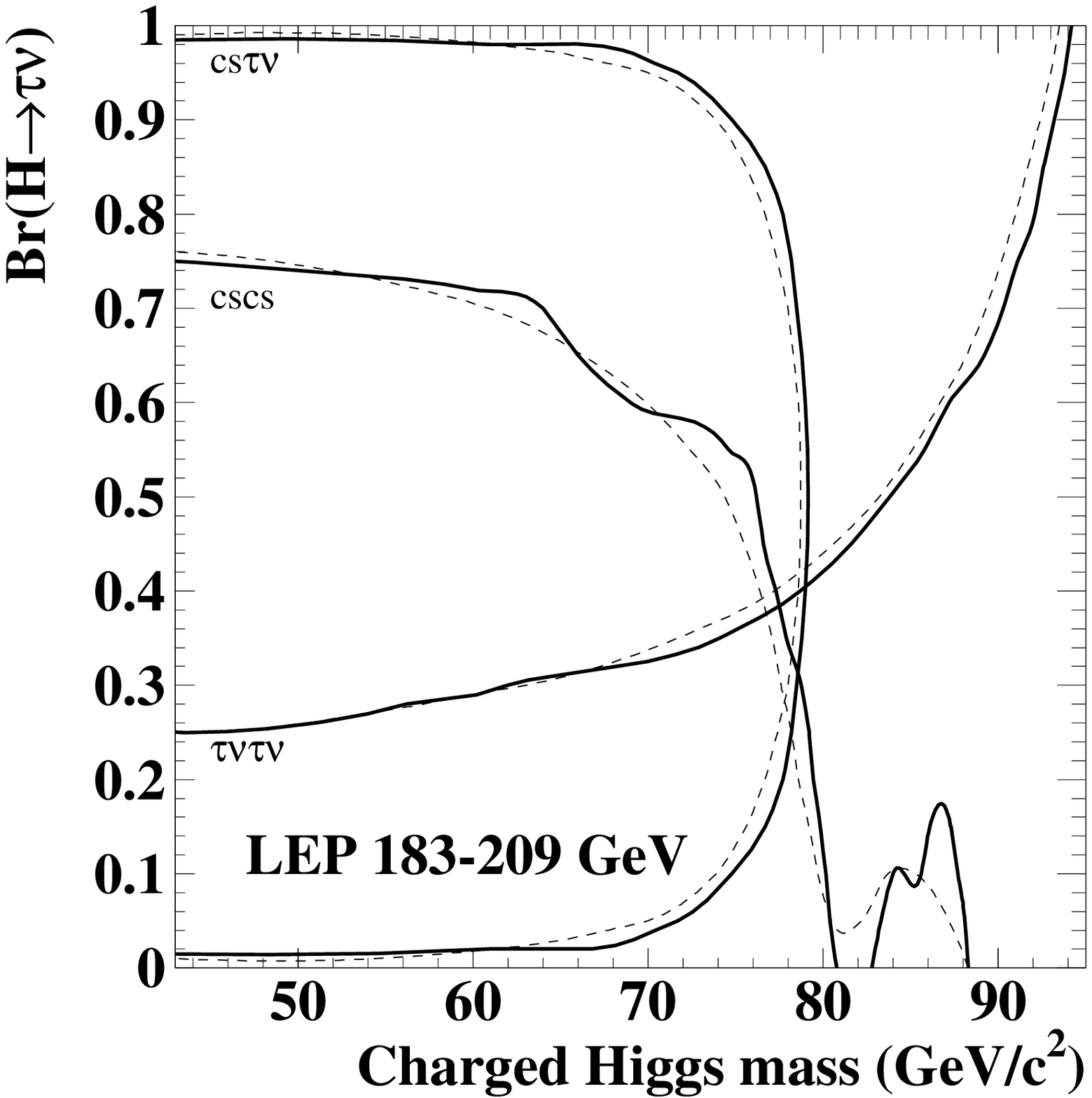}
\caption[]{\small \it Type II 2HDM:
regions in the Br(\Hp\ra\tp$\nu$) vs \mHpm\ plane excluded at the 95\% or 
higher C.L., based on the combined data collected by the four LEP experiments 
at centre-of-mass 
energies from 183 to 209~GeV, for each of the three decay channels separately.
The solid (dashed) lines are the observed (expected) limits.}
\label{charged-limchan}
\end{center}
\end{figure}

The combined results for the Type~II 2HDM are summarised in 
Figure~\ref{charged-limit}, 
which shows the expected median and observed mass limits, while 
the contribution of each of the three decay channels to the overall limit 
is presented in Figure~\ref{charged-limchan}. It is worth noting that:
\begin{itemize}
\item the purely leptonic channel alone excludes charged Higgs masses 
above the W mass, down to Br(\Hp\ra\tp$\nu$)
around 0.45. In this channel, the mass of the Higgs
boson cannot be reconstructed, due to the 
presence of two neutrinos in the final state. As a consequence, the W 
boson pair
background is diluted and the analysis is sensitive up to $\sqrt{s}$/2.     
The limit drops rapidly for 
Br(\Hp\ra\tp$\nu$) below 0.45, due to a rapid decrease of the signal rate 
in this final state;
\item the mixed channel alone cannot exclude charged Higgs mass values 
up to the 
W mass, even when it contributes maximally, for Br(\Hp\ra\tp$\nu$) = 0.5. 
For this value, the observed limit 
is only slightly above 79~\Gcs, due to the large \ee\ra\WW\ 
background. This channel has the
best coverage in terms of Br(\Hp\ra\tp$\nu$), as shown in Figure 3;
\item the hadronic channel is the most difficult one; for masses close to the W
mass, the sensitivity is reduced due to the large \ee\ra\WW\ background.
The sensitivity at higher masses is improved (a gain of 10 \Gcs\ on the 
expected limit), and the observed limit as well  
(note the excluded ``island'' 
at Br(\Hp\ra\tp$\nu$) close to zero) with respect to the results of 
individual experiments; 
\item the difference between the expected and observed limit seen 
in Figure~\ref{charged-limit} 
for Br(\Hp\ra\tp$\nu$) from 0.35 to 0.85 results from the
excess of observed events already mentioned (see Figure~\ref{charged-clb}) 
in the mixed channel above \mHpm\ = 84 \Gcs.
\end{itemize}

The combined 95\% C.L. \mHpm\ lower limits are listed in 
Table~\ref{table-ch-limits} for Br(\Hp\ra\tp$\nu$)=0, 1, together with the 
limit that is independent of the fermionic decay mode. Taking the lowest 
of the observed limits from Table~\ref{table-ch-limits},
we quote a 95\% C.L. lower bound of 80~\Gcs\  for the
mass of the charged Higgs boson in type II 2HDM under the assumption of
pure fermionic decays of the charged Higgs boson.
Thus the hypothesis of a charged Higgs boson degenerate in mass with the
W boson is not excluded at the 95\% confidence level with LEP data.
The limits around the W mass are very sensitive to the modelling of the
W pairs background. Taking the uncertainties in the background modelling
(including the uncertainty due to different simulated W masses)
into account results in a downward shift of these limits by 600 and 500~\Mcs\ 
for Br(\Hp\ra\tp$\nu$)=0 and 0.5, respectively. 

\begin{table} [hbtp]
\begin{center}
\caption{\small\it The combined 95\% C.L. lower bounds on the mass of the
charged Higgs boson (in \Gcs), expected and observed, for fixed 
values of the branching ratio Br(\Hp\ra\tp$\nu$)
and for any Br(\Hp\ra\tp$\nu$). All mass limits have been rounded down to the
nearest half a \Gcs\ to take into account the effect of neglecting the
correlations between systematic uncertainties.
(*) The interval from 83 to 88 \Gcs\ is also excluded at the 95\% C.L.}
\label{table-ch-limits}
\begin{tabular}{lcc}
\hline\noalign{\smallskip}
       & Expected limit (median) & Observed limit \\
\noalign{\smallskip}\hline
Br(\Hp\ra\tp$\nu$)=0  & 88   & 80.5 (*) \\
Br(\Hp\ra\tp$\nu$)=1  & 93.5 & 94       \\
Any Br(\Hp\ra\tp$\nu$)& 79.5 & 80       \\
\noalign{\smallskip}\hline
\end{tabular}
\end{center}
\end{table}

\begin{figure}[htb]
\begin{center}
\begin{tabular}{cc}
\includegraphics[width=0.45\columnwidth]{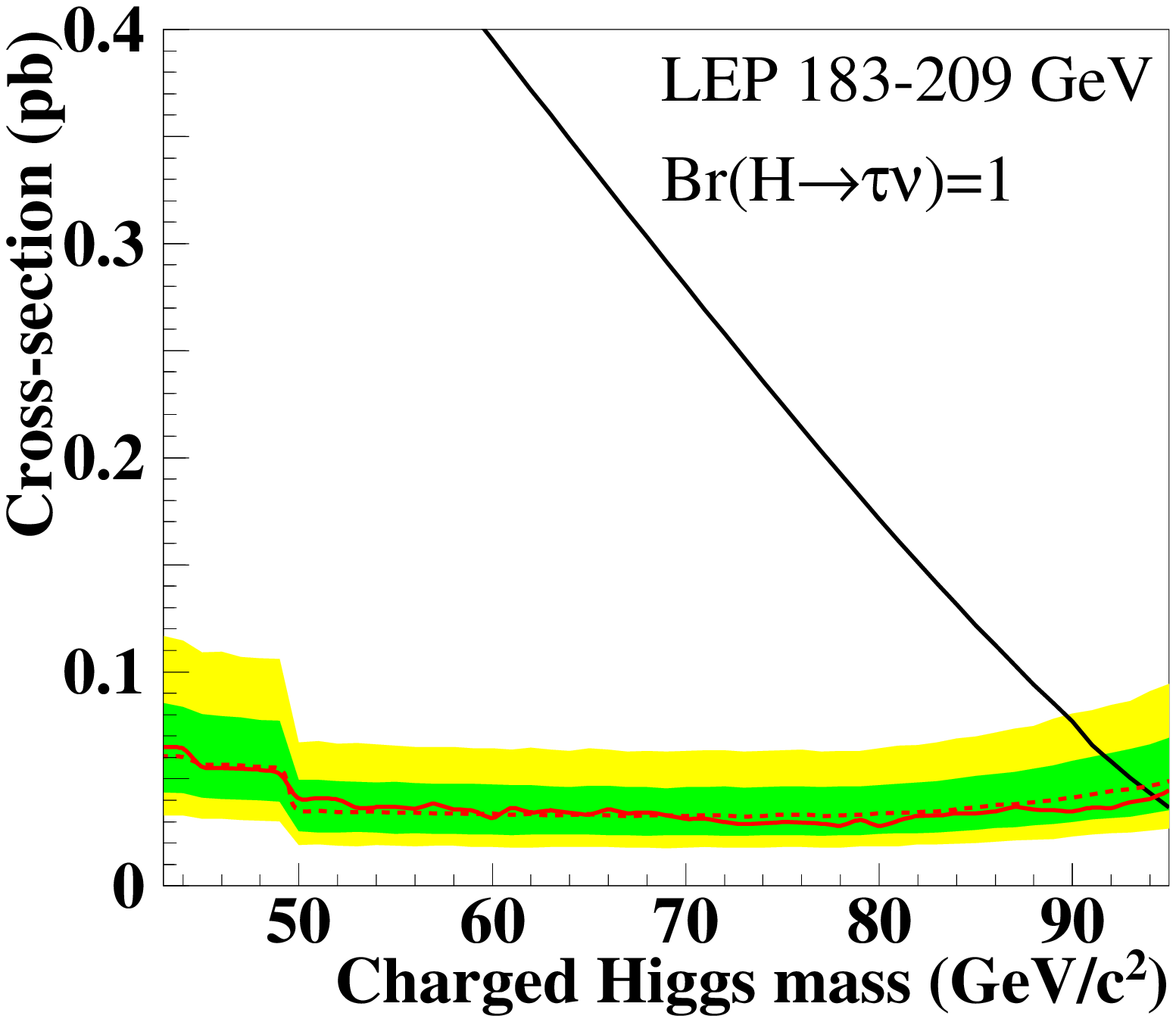}    &
\includegraphics[width=0.45\columnwidth]{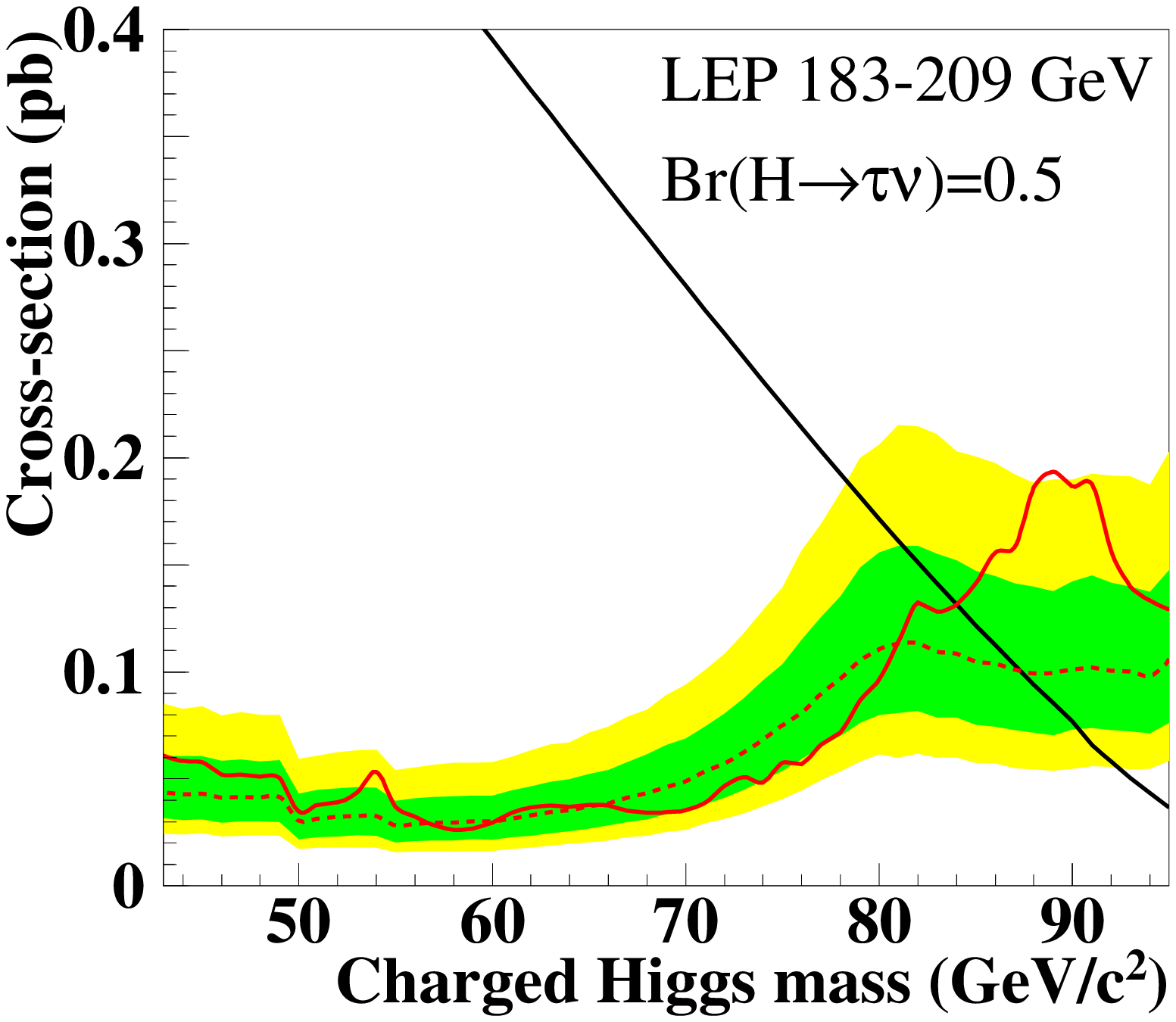}   \\
\includegraphics[width=0.45\columnwidth]{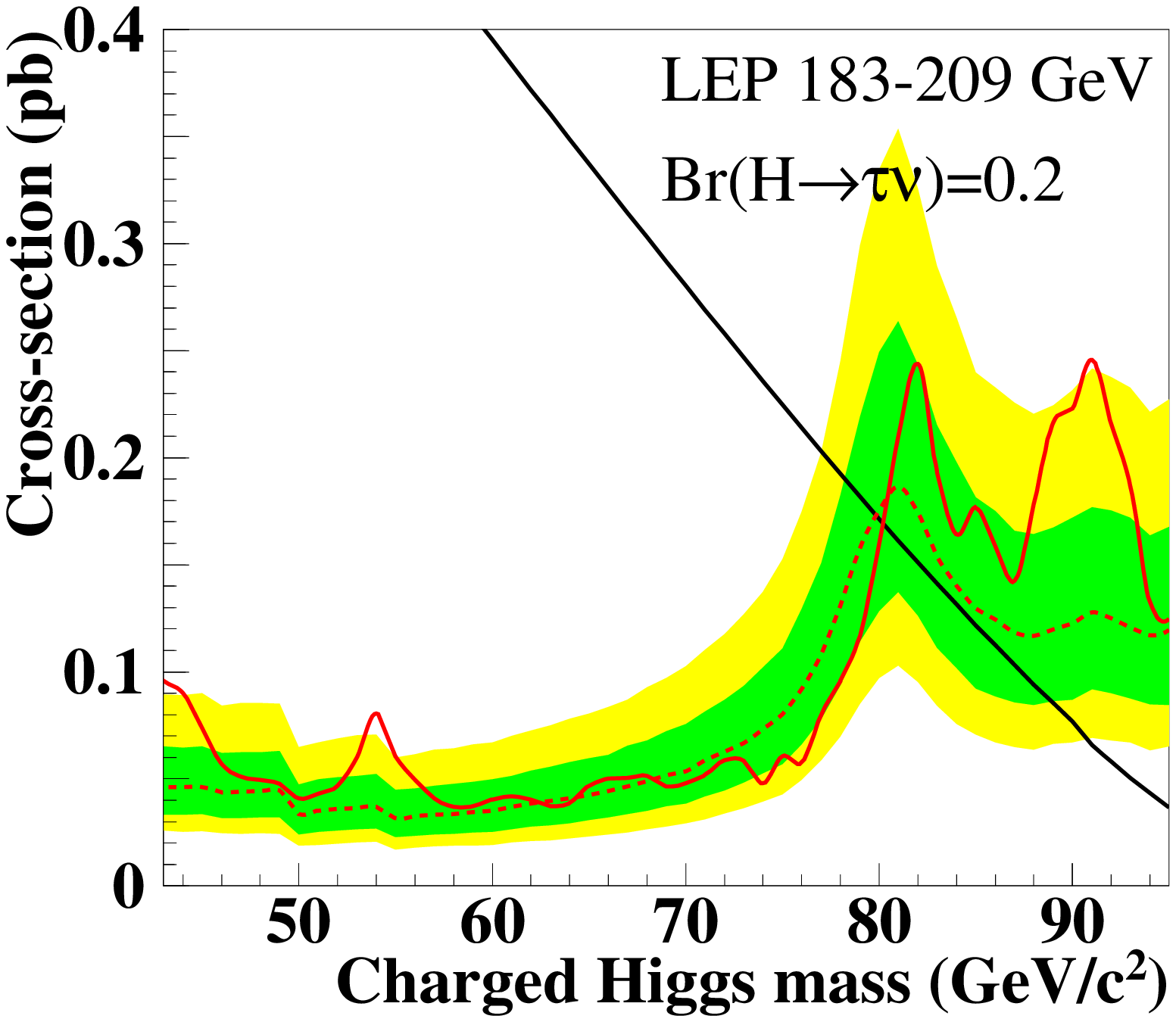}    &
\includegraphics[width=0.45\columnwidth]{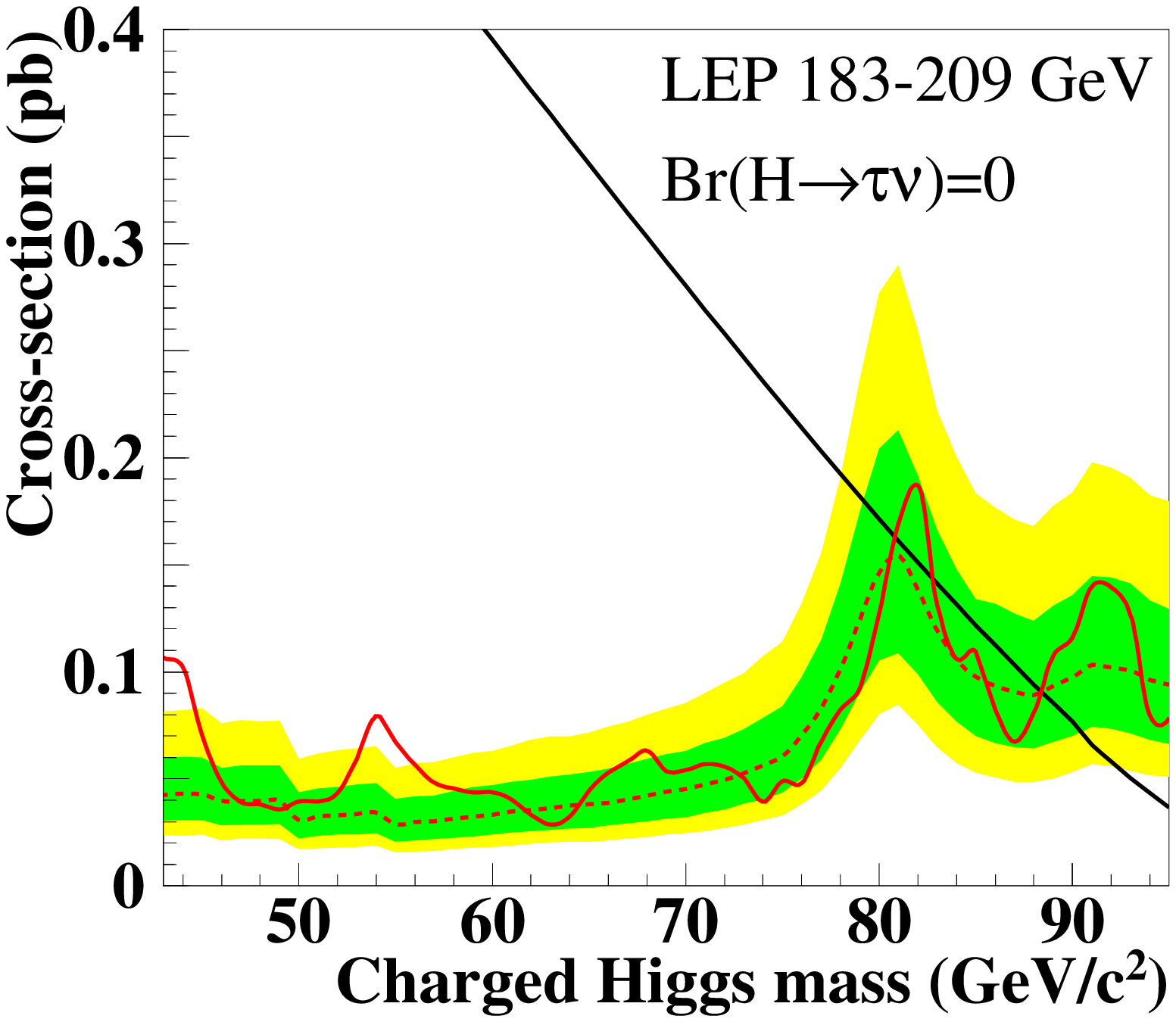}   \\
\end{tabular}
\caption[]{\small \it Type II 2HDM:
the 95\% C.L. upper limits on the production cross-section as a function of 
\mHpm\  for four different values of the branching ratio Br(\Hp\ra\tp$\nu$), 
combining the data collected by the four LEP experiments at centre-of-mass 
energies from 183 to 209~GeV. The solid lines represent the observed exclusion
limits, while
the expected exclusion limits are indicated by the dashed lines. 
The shaded bands represent the $\pm 1\sigma$ and $\pm 2\sigma$ excursions
around the expected limits. The intersections of the curves (solid or dashed) 
with the thick line showing the theoretical (tree-level) charged Higgs
cross-section represent the (observed or expected) 95\% C.L. lower limits 
on the charged Higgs boson mass.}
\label{xsec-limit}
\end{center}
\end{figure}

Figure~\ref{xsec-limit} shows the 95\% C.L. upper bound on 
the \ee\ra\Hp\Hm\ cross-section (with $\pm$ 1$\sigma$ and $\pm$ 2$\sigma$ 
bands) for four values of Br(\Hp\ra\tp$\nu$), namely 1, 0.5, 0.2  
(which corresponds to the weakest limit) and 0.
The thick black curve is the 2HDM tree-level prediction for that cross-section.

\clearpage

\section{Combined searches in 2HDM of type I}
\label{sec:3}
\begin{figure}[htb]
\begin{center}
\includegraphics[width=0.95\columnwidth]{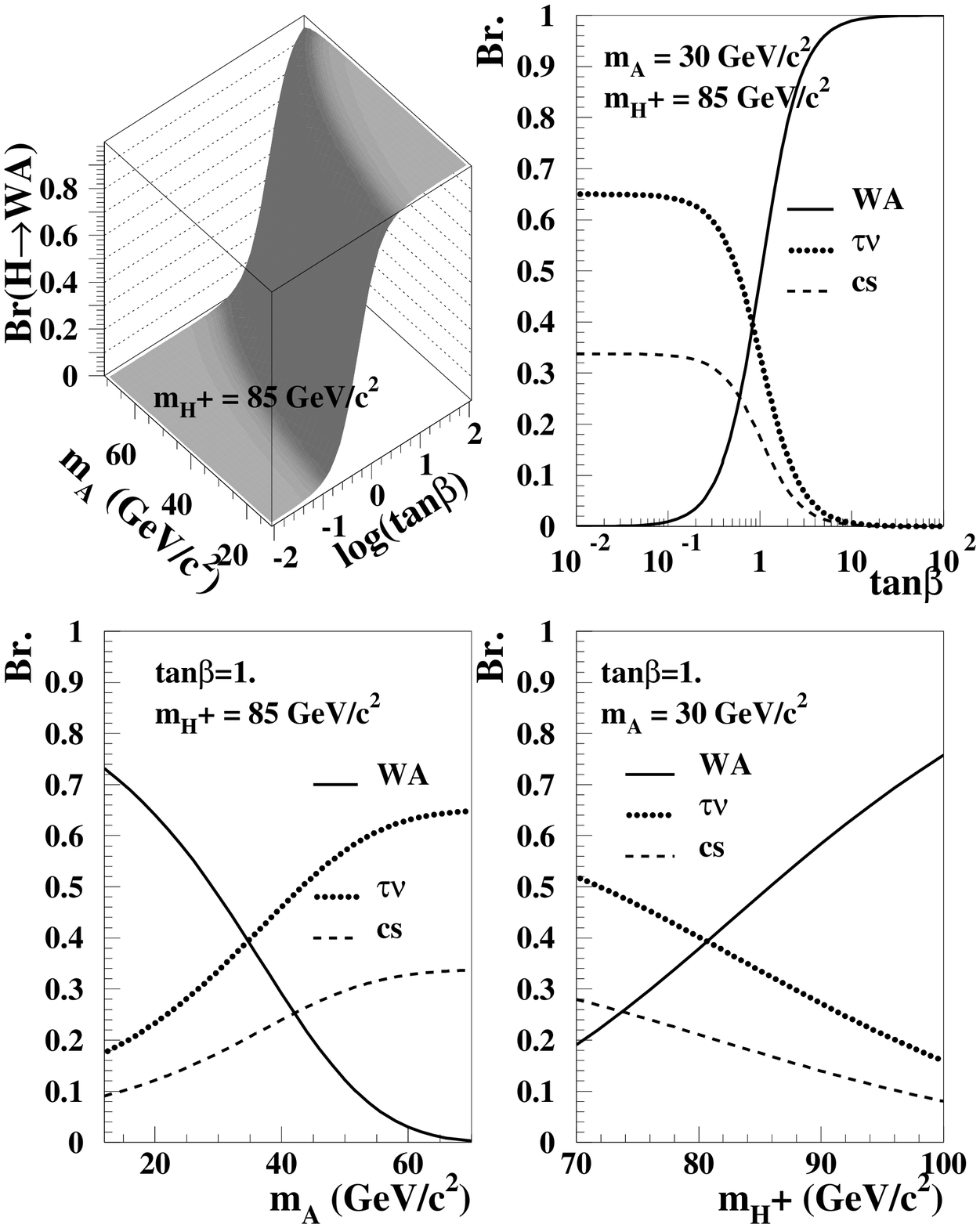}
\caption[]{\small \it Type I 2HDM: decay branching fractions as 
functions of the boson masses and \tanb.}
\label{fig:type1}
\end{center}
\end{figure}

An alternative set of models, type I 2HDMs, assume that all 
fermions couple to the same Higgs doublet. In this case all fermions couple 
proportionally to 1/\tanb\ to the charged Higgs boson and fermionic decays 
are suppressed for medium to large \tanb\ values. Consequently, if a neutral 
Higgs boson $\Phi$ (representing either A or the lightest CP-even scalar h) 
is sufficiently light, the decay to $\wn$ can be dominant even in the 
range of charged Higgs masses of interest
at LEP (where $\mathrm{W}^*$ indicates an off-shell W boson).
While searches for a CP-even neutral Higgs boson exclude such a particle 
for masses below  82 \Gcs\ independently of its decay~\cite{opalindep}, the 
existence of a light CP-odd neutral Higgs boson, A, is not excluded by 
experiment~\cite{delphi2hdm}.
Hence, the search for the process \Hpm\ra$\wa$ is fully justified.
Figure~\ref{fig:type1} shows the predicted branching ratios of the charged
Higgs bosons for various choices of parameters of type I models. 
For all kinematically allowed values of the A mass, \mA,
the possible charged Higgs boson decays are predominantly fermionic for 
low \tanb\ and predominantly bosonic for high \tanb . 
Between these two extreme cases, the branching ratios change rapidly as a
function of \tanb\
(between typically 0.1 and 10) and slower as a function of \mA, appearing 
earlier
in \tanb\ for lower \mA. The ratio between the two competing
fermionic decays ($\tau \nu$ over \csbar) is almost independent of the charged 
Higgs boson mass (see lower part of the figure), as expected from the
Yukawa coupling which only depends upon the masses involved. 

To cover the possibility of a light A boson the final states $\HWAWA$ 
and $\HWATN$ were also searched for by DELPHI~\cite{delphi} and 
OPAL~\cite{opal}. The channel $\HWACS$ was not considered because its 
contribution is expected to be small for all $\tan\beta$. The A
boson was searched for through its decay into two b-jets, restricting the 
A mass to be above 12 \Gcs. Type I models are explored through the 
combination of all five decay channels, namely the final 
states \csbar\cbars, \csbar$\tau\nu$, \tp$\nu$\tm\nubar,
 $\HWAWA$ and $\HWATN$ (and their charge conjugates). The combination of the 
experimental search results is performed for branching ratio values predicted 
by the model as a function of \tanb\ and \mA. Where there was a 
possible overlap between two search channels, the one providing less expected 
sensitivity was ignored to avoid double counting. This is the case in the 
intermediate region in \tanb\ for purely hadronic channels 
($\HWAWA$ and \csbar\cbars) on the one hand and the semi-leptonic 
channels ($\HWATN$ and \csbar\tm\nubar) on the other. A three-dimensional 
scan was performed with the following ranges and steps: \mHpm\ from 43 to 
95 \Gcs\ in 1 \Gcs\ steps, \mA\ covering 12 \Gcs, then 15 to 75 \Gcs\ in 
5 \Gcs\ steps, and \tanb\ from 0.1 to 100 in steps of 0.2 in $\log(\tan\beta)$.

\begin{figure}[h]
\vspace{-.4cm}
\begin{center}
\begin{tabular}{cc}
\includegraphics[width=0.45\columnwidth]{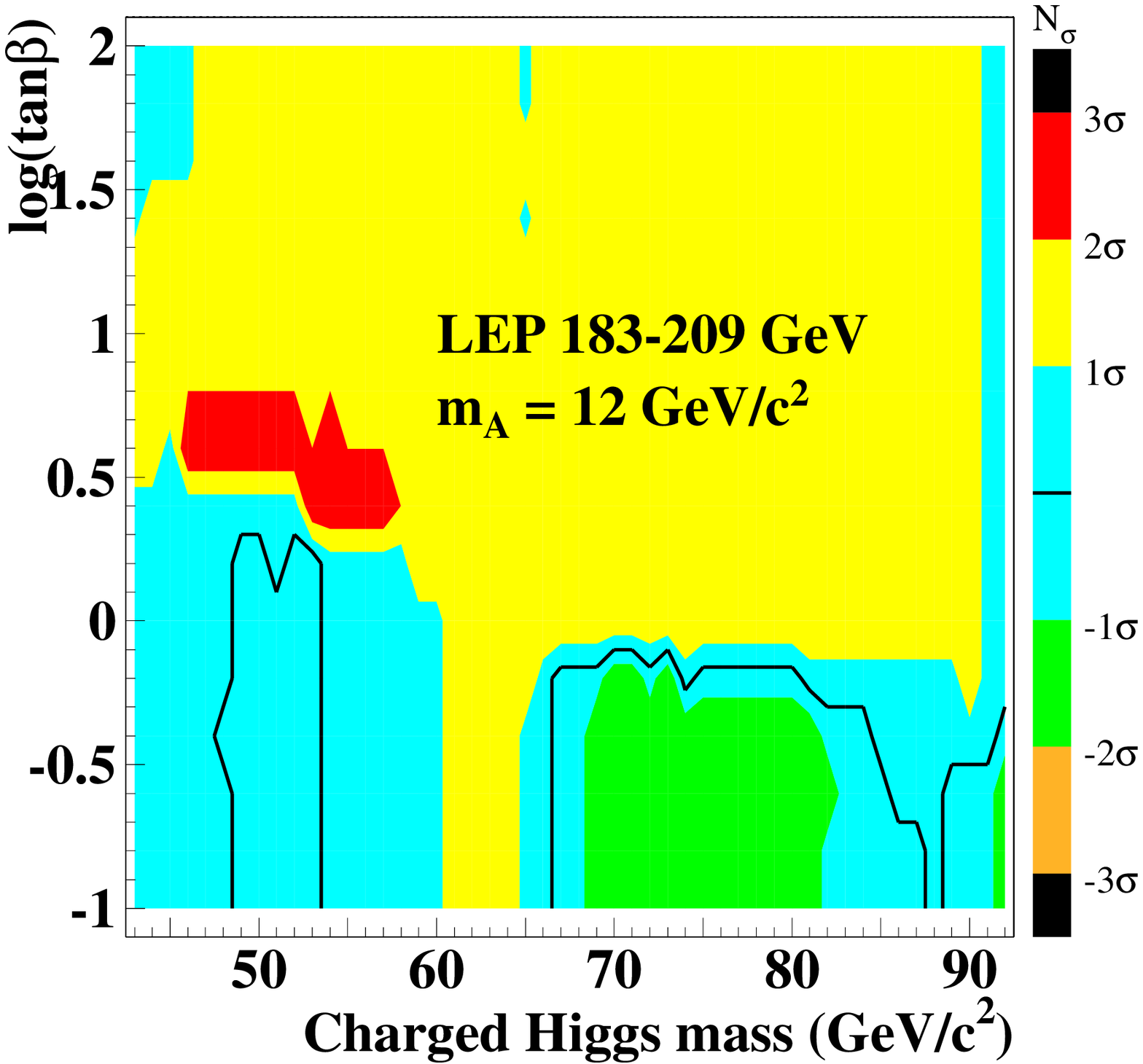}  &
\includegraphics[width=0.45\columnwidth]{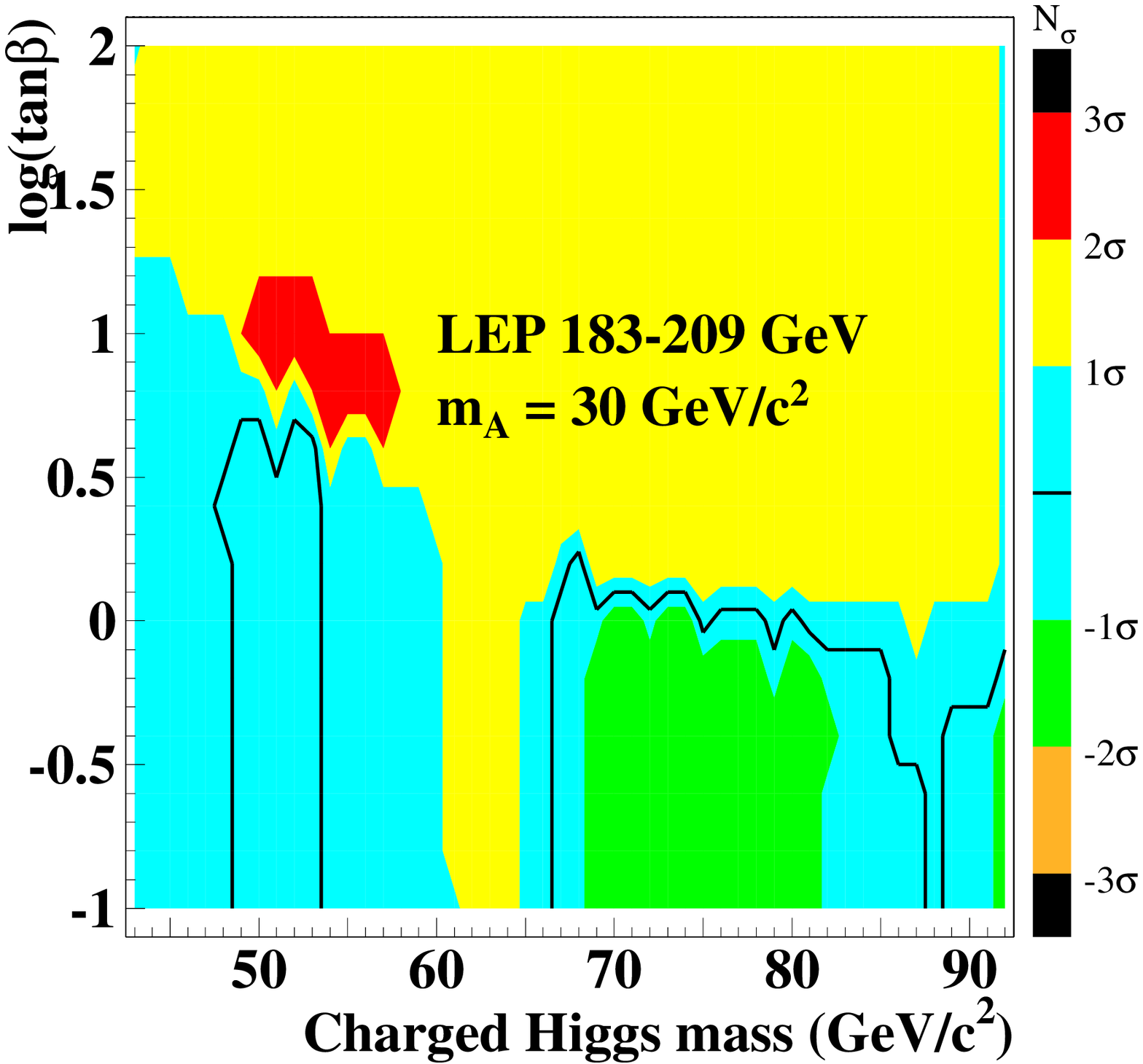}  \\
\includegraphics[width=0.45\columnwidth]{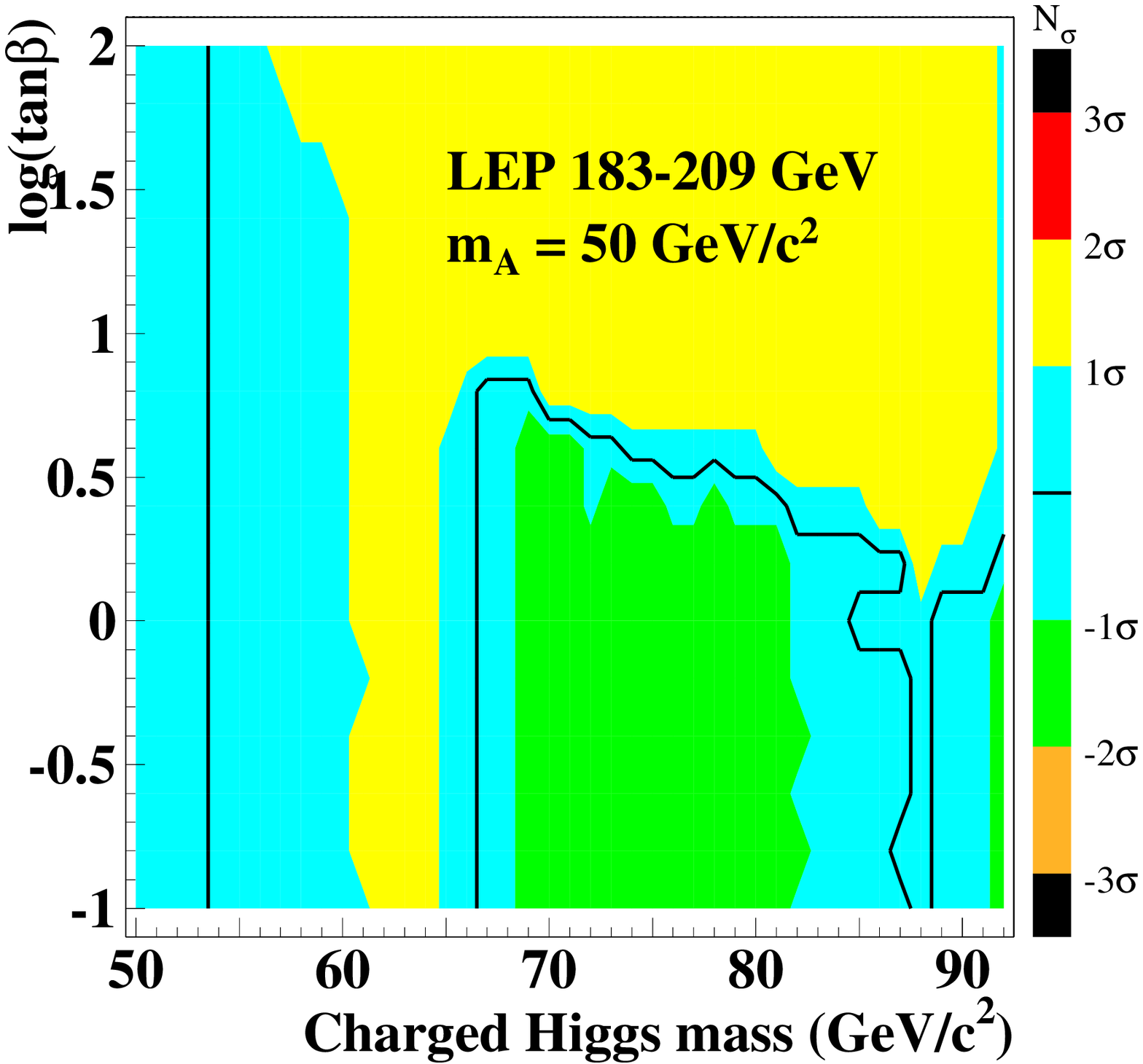}  &
\includegraphics[width=0.45\columnwidth]{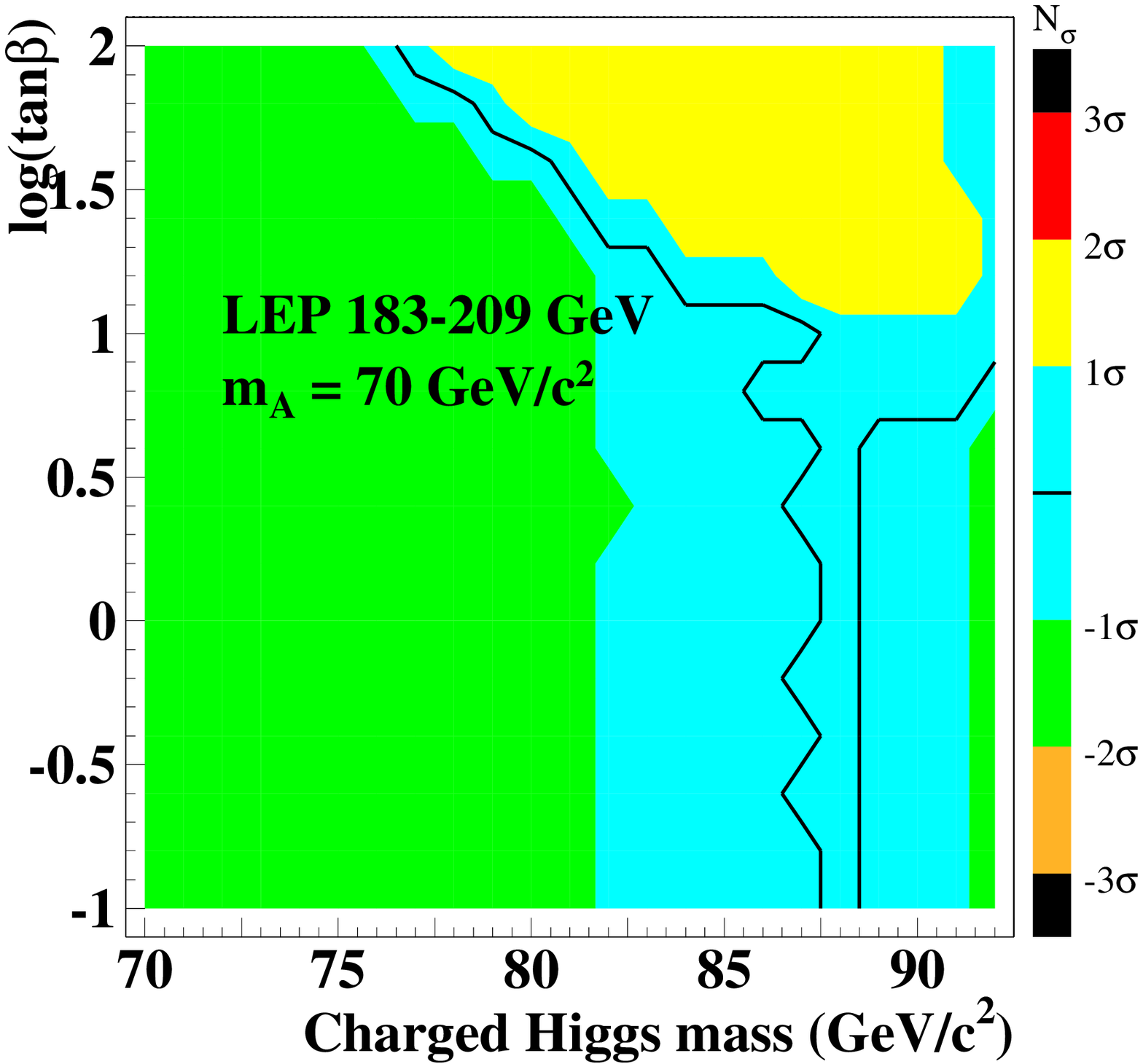}  \\
\includegraphics[width=0.45\columnwidth]{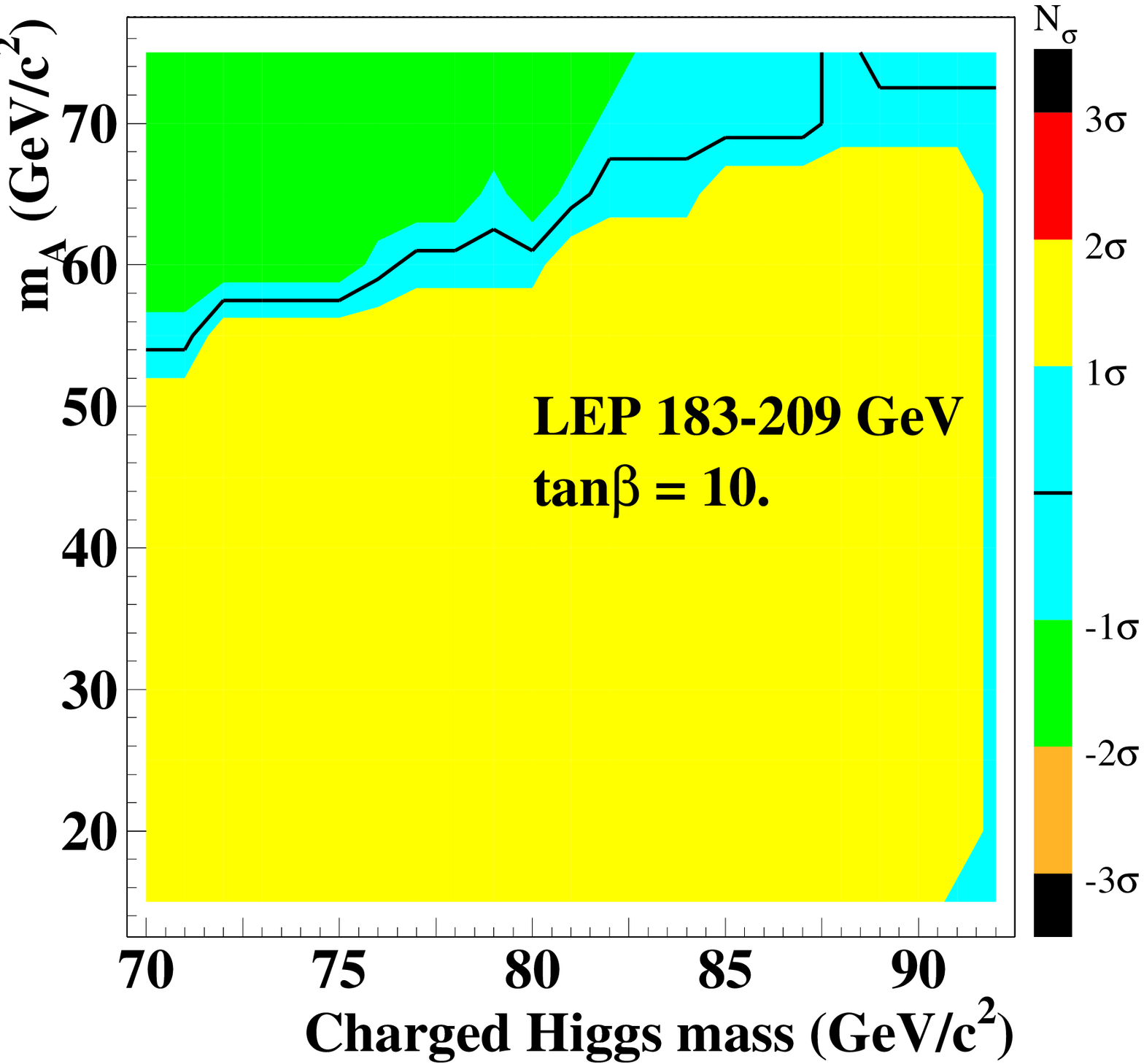}  &
\includegraphics[width=0.45\columnwidth]{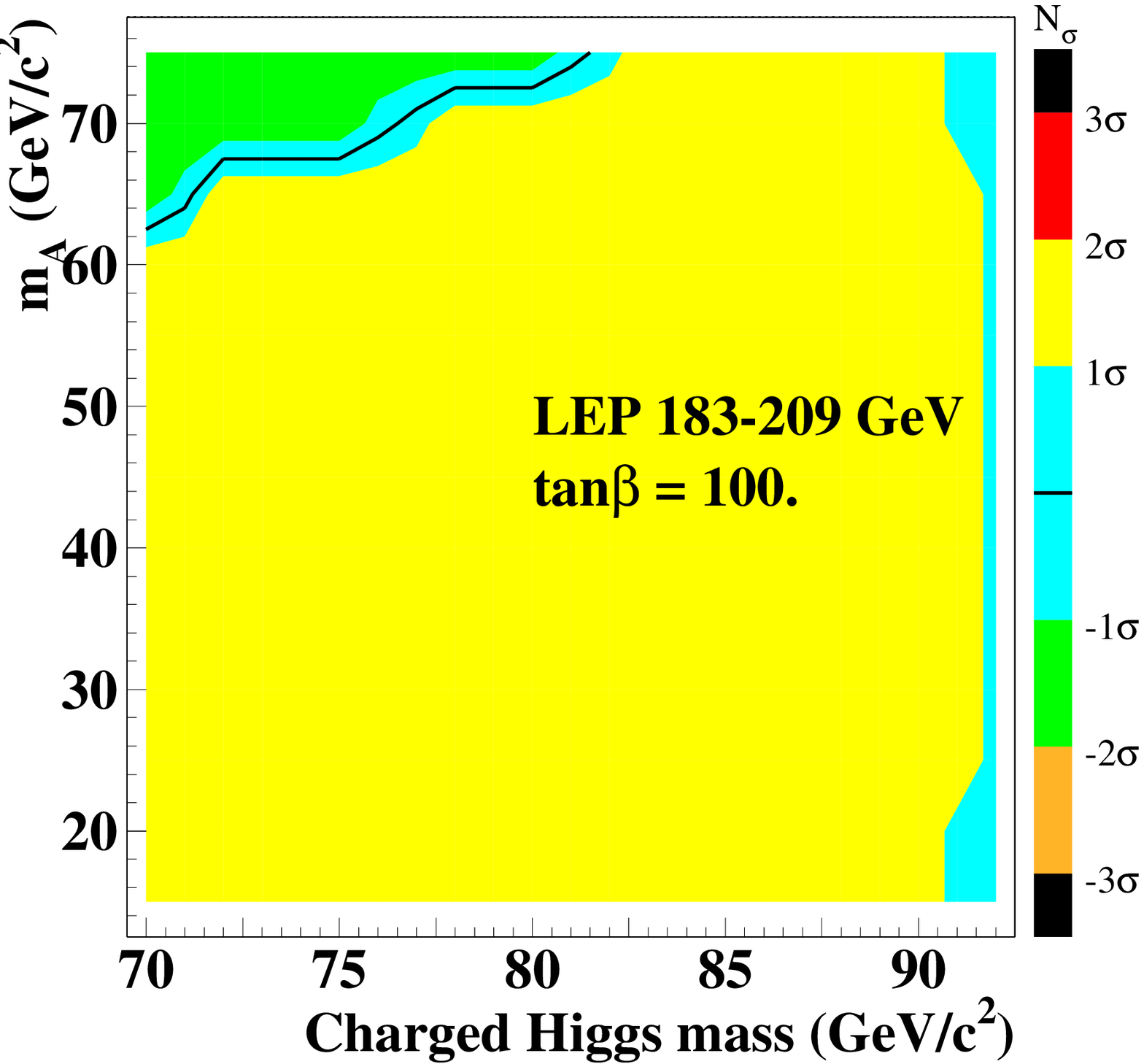}  \\
\end{tabular}
\end{center}
\caption{\small \it Type I 2HDM: 
contours based on the observed p-values $CL_b$
as a function of $m_{\mathrm{H}^{\pm}}$ and $\tan \beta$ or $m_{\mathrm{A}}$,
indicating the statistical significance, N$_{\sigma}$,
of local departures from the background expectation, for four
values of $m_{\mathrm{A}}$ and two values of $\tan \beta$.
The black solid line indicates the change of sign of this significance, 
i.e. where there is a transition from excess to deficit.}  
\label{type1clb}
\end{figure}

\begin{figure}[h]
\includegraphics[width=0.99\columnwidth]{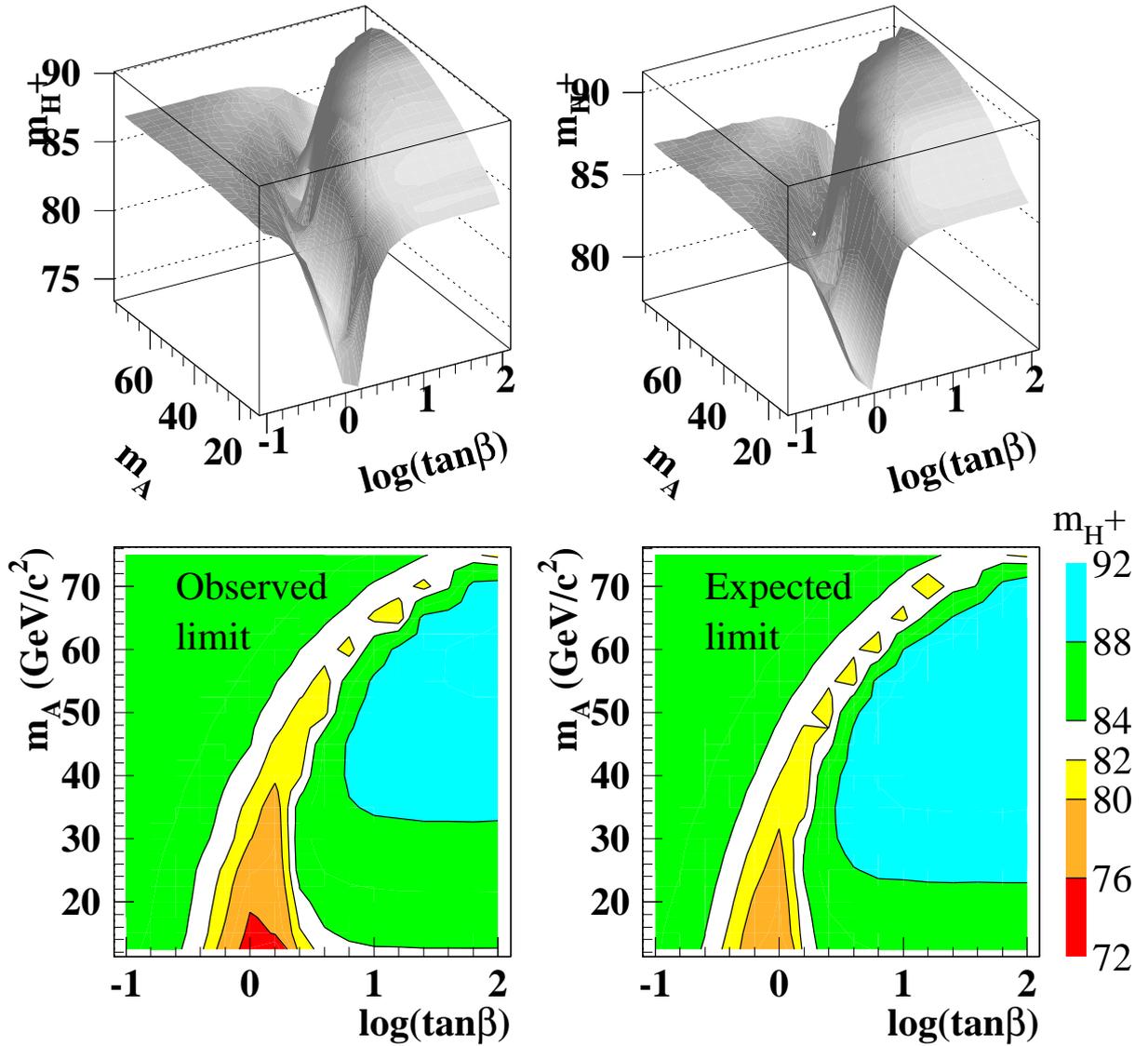}
\caption{\small \it Type I 2HDM:
Observed (left) and expected (right) 95\% C.L. lower limits 
on the mass of the charged Higgs boson. The colour-mass correspondence is
indicated on the right hand side (units are \Gcs).}
\label{type1lim}
\end{figure}

Figure~\ref{type1clb} shows the observed $CL_b$, for four values 
of \mA\ and two values of $\tan \beta$. A slight excess for low and 
intermediate A masses in the high $\tan \beta$ region where 
the bosonic decays dominate is observed, resulting in observed limits 
generally weaker than expected (see Figure~\ref{type1lim}). Three main 
features are visible in Figure~\ref{type1lim}, 
two plateaux and a valley between them:
\begin{itemize}
\item the first plateau, at low $\tan \beta$, corresponds to the case when
the fermionic channels dominate. Both expected and observed 
limits are above 86 \Gcs;
\item the valley is somewhat of an artefact. It is due to the conservative 
approach of considering only the most sensitive channel when two overlapping
channels contribute. The difference between
expected and observed mass limits reaches 4.5 \Gcs\ in the extreme case (when 
$\tan \beta$ = 1.6 and \mA\ = 12 \Gcs);
\item the second plateau, at high $\tan \beta$, corresponds to the case 
when the
bosonic channels dominate. The small excess seen in Figure~\ref{type1clb} 
corresponds to a small difference between expected and observed 
charged Higgs mass limits, which is always less than 2.2 \Gcs.
\end{itemize}

\begin{figure}[h]
\includegraphics[width=0.99\columnwidth]{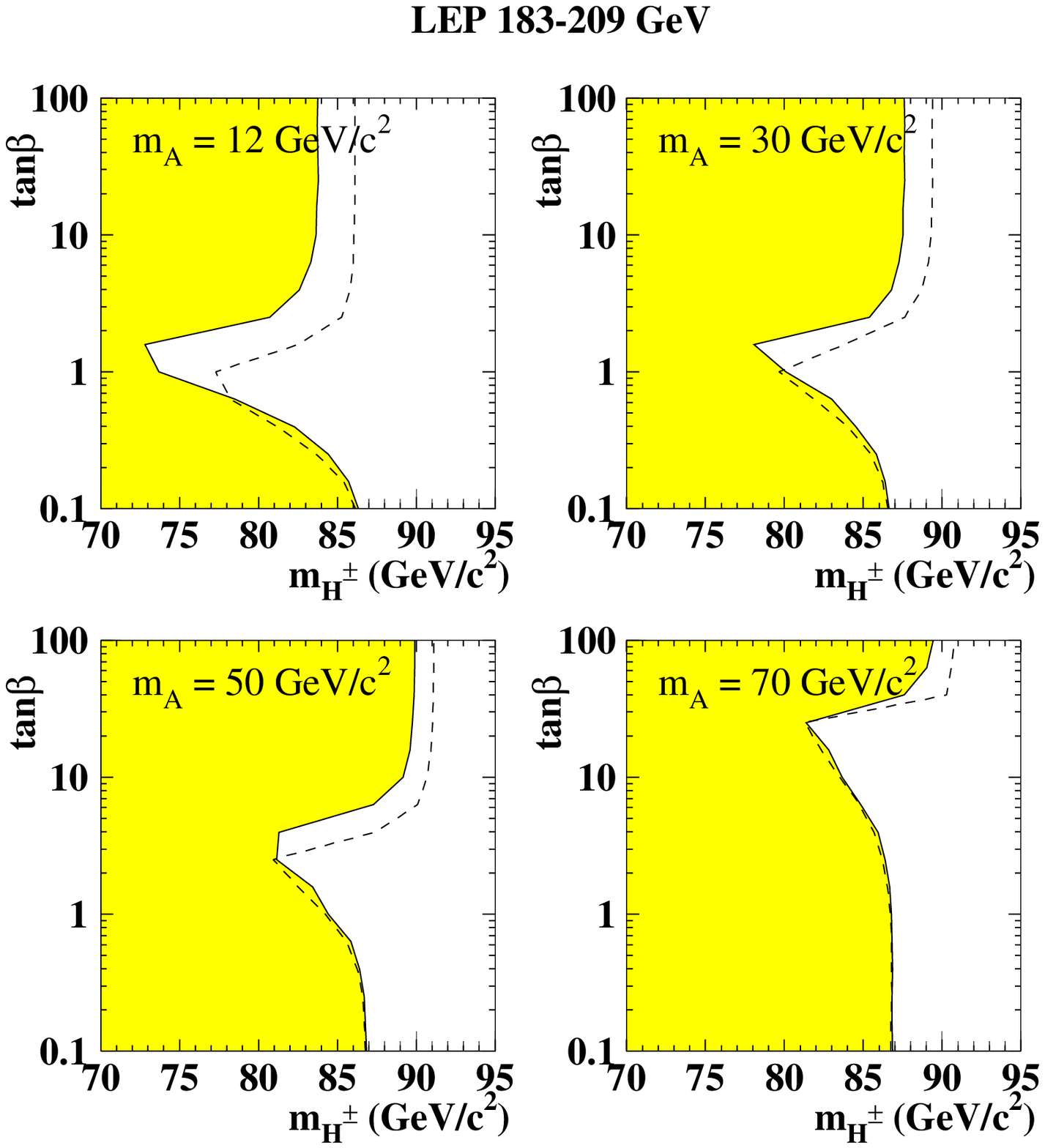}
\caption{\small \it Type I 2HDM: Excluded regions at 95\% C.L. in the 
(\mHpm,\tanb) plane for different values of \mA: 12, 30, 50 and 70 \Gcs. 
The dashed line represents the expected exclusion limit.}
\label{globma}
\end{figure}

Figure~\ref{globma} shows the excluded regions at the 95\% C.L.
in the plane (\mHpm,\tanb) for four values of \mA,
namely 12, 30, 50 and 70 \Gcs, together with the expected exclusion limits.

Table~\ref{type1limits} summarizes these results. For low 
$\tan \beta$ ($\tan \beta$ below 0.5)
where the bosonic contribution is vanishingly small, the \mHpm\ lower limits 
(above 86 \Gcs) are almost independent of \mA.                
On the other hand, for high
$\tan \beta$ (equal to or greater than 10) where the bosonic 
channels dominate, the sensitivity is maximal for intermediate A masses 
(\mA\ around 50 \Gcs). Outside the valley, the limit is 
always above 84 \Gcs . Finally, the lowest limits always correspond to
the cases in the valley, thus depending both on $\tan \beta$
and \mA . The lowest (observed) limit is 72.5 \Gcs , for
$\tan \beta$ = 1.6 and \mA\ =12 \Gcs . This limit rises to 76.5 \Gcs\ for  
\mA\ =20 \Gcs\ and the difference between expectation and observation is 
reduced to 1 \Gcs .

\begin{table*} [hbtp]
\begin{center}
\caption{\small\it Observed lower limits on the charged Higgs mass in \Gcs\ at
95\% C.L. for different values of \mA\ (in \Gcs) and $\tan \beta$. The expected
median limits are shown in parentheses. The last column (last row)
show the weakest limit for a fixed $\mathrm{A}$ mass and any $\tan \beta$ 
(for a fixed $\tan \beta$ and any $\mathrm{A}$ mass). The mass limits have 
been rounded down to the nearest half a \Gcs to take into account 
the effect of neglecting correlations between systematic uncertainties.}
\label{type1limits}
\begin{tabular}{c|ccccc}
\hline\noalign{\smallskip}
\mA  & \tanb\ = 0.1 &\tanb = 1 & \tanb = 10 &  \tanb = 100 &  minimum  \\
\noalign{\smallskip}\hline
12 & 86.0 (86.0) & 73.5 (77.0) & 83.5 (86.0) & 84.0 (86.0) & 72.5 (77.0) \\
20 & 86.5 (86.0) & 76.5 (77.5) & 85.5 (87.0) & 85.5 (87.0) & 76.5 (77.5) \\
30 & 86.5 (86.5) & 80.0 (79.5) & 87.5 (89.0) & 87.5 (89.0) & 78.0 (79.5) \\
50 & 86.5 (86.5) & 84.0 (84.0) & 89.0 (90.0) & 89.5 (91.0) & 81.0 (80.5) \\
70 & 86.5 (86.5) & 86.5 (86.5) & 83.5 (83.5) & 89.0 (90.5) & 81.0 (81.0) \\
\noalign{\smallskip}\hline
minimum & 86.0 (86.0) & 73.5 (77.0) & 81.5 (81.0) & 81.0 (81.0) & 72.5 (77.0)\\
\noalign{\smallskip}\hline
\end{tabular}
\end{center}
\end{table*}

\section{Summary}

The results of the searches carried out by the four LEP experiments for 
charged Higgs bosons have been statistically combined and interpreted in
2HDMs. No significant excess over the SM 
background is observed, and the exclusion limits are extended by
several \Gcs\ with respect to the final results of the individual
collaborations~[3,4,6,7] and the previous combination~\cite{previous}.
In the type II 2HDM scenario,
assuming that the two decays \Hp\ra\csbar\ and \Hp\ra\tp$\nu$ exhaust 
the \Hp\ decay width, mass limits are obtained as a function of the 
branching ratio Br(\Hp\ra$\tau^+\nu$). A 95\% C.L. lower limit on the charged 
Higgs mass, independent of its fermionic decay modes, is found to be 80~\Gcs.
Thanks to analyses by DELPHI and OPAL in the bosonic $\wa$ decay channels,
a new scenario, for type I 2HDM, is also studied. In this case, masses 
of the charged Higgs boson below 72.5~\Gcs\ are excluded at the 95\% C.L.
for A masses above 12 \Gcs .

\section*{Acknowledgements}
We would like to thank the CERN accelerator division for the excellent
performance of the LEP accelerator in its high-energy phase. The LEP working 
group for Higgs boson searches would also
like to thank the members of the four LEP experiments for providing their
results and for valuable discussions concerning their combination.


\begin{thebibliography}{99}
\bibitem{lhwg}
  The LEP Working Group for Higgs Boson 
  Searches consists of members of the four LEP collaborations and 
  of theorists among whom S.~Heinemeyer is co-author of this paper.
\end{thebibliography}
\end{document}